\RequirePackage{lineno}
\documentclass[aps,prc,reprint,floatfix,showpacs,showkeys,superscriptaddress]{revtex4-2}
\usepackage{soul}
\setstcolor{red}
\usepackage{graphicx}
\usepackage{tikz}
\usepackage{verbatim}
\usepackage{color}
\usepackage{amsmath}

\begin{document}
\title{New Measurements of the Deuteron to Proton $F_2$ Structure Function Ratio}
\date{\today}
\newcommand*{\MSU }{Mississippi State University, Mississippi State, Mississippi 39762, USA}
\newcommand*{\UVA }{University of Virginia, Charlottesville, Virginia 22903, USA}
\newcommand*{\JLAB }{Thomas Jefferson National Accelerator Facility, Newport News, Virginia 23606, USA}
\newcommand*{\REG }{University of Regina, Regina, Saskatchewan S4S 0A2, Canada}
\newcommand*{\NCAT }{North Carolina A \& T State University, Greensboro, North Carolina 27411, USA}
\newcommand*{\KENT }{Kent State University, Kent, Ohio 44240, USA}
\newcommand*{\ZAG }{University of Zagreb, Zagreb, Croatia}
\newcommand*{\TEMP }{Temple University, Philadelphia, Pennsylvania 19122, USA}
\newcommand*{\YER }{A.I. Alikhanyan  National  Science  Laboratory \\ (Yerevan  Physics
Institute),  Yerevan  0036,  Armenia}
\newcommand*{\WM }{The College of William \& Mary, Williamsburg, Virginia 23185, USA}
\newcommand*{\CUA }{Catholic University of America, Washington, DC 20064, USA}
\newcommand*{\HU }{Hampton University, Hampton, Virginia 23669, USA}
\newcommand*{\FIU }{Florida International University, University Park, Florida 33199, USA}
\newcommand*{\CNU }{Christopher Newport University, Newport News, Virginia 23606, USA}
\newcommand*{\JAZ }{Jazan University, Jazan 45142, Saudi Arabia}
\newcommand*{\UTENN }{University of Tennessee, Knoxville, Tennessee 37996, USA}
\newcommand*{\OHIO }{Ohio University, Athens, Ohio 45701, USA}
\newcommand*{\UCONN }{University of Connecticut, Storrs, Connecticut 06269, USA}
\newcommand*{\SBU }{Stony Brook University, Stony Brook, New York 11794, USA}
\newcommand*{\ODU }{Old Dominion University, Norfolk, Virginia 23529, USA}
\newcommand*{\ANL }{Argonne National Laboratory, Lemont, Illinois 60439, USA}
\newcommand*{\BOULDER }{University of Colorado Boulder, Boulder, Colorado 80309, USA}
\newcommand*{\ORSAY }{Institut de Physique Nucleaire, Orsay, France}
\newcommand*{\UNH }{University of New Hampshire, Durham, New Hampshire 03824, USA}
\newcommand*{\JMU }{James Madison University, Harrisonburg, Virginia 22807, USA}
\newcommand*{\RUTG }{Rutgers University, New Brunswick, New Jersey 08854, USA}
\newcommand*{\CMU }{Carnegie Mellon University, Pittsburgh, Pennsylvania 15213, USA}
\newcommand*{\KSU}{King Saud University, Riyahd 11451, Saudi Arabia}
\newcommand{\VMI}{Virginia Military Institute, Lexington, Virginia, 24450 USA}

\author{D.~Biswas}\affiliation{\HU}   
\author{F.~A.~Gonzalez}\affiliation{\SBU}            
\author{W.~Henry}\affiliation{\JLAB}            
\author{A.~Karki}\affiliation{\MSU}            
\author{C.~Morean}\affiliation{\UTENN}            
\author{A.~Nadeeshani}\affiliation{\HU}   
\author{A.~Sun}\affiliation{\CMU}               

\author{D.~Abrams}\affiliation{\UVA} 
\author{Z.~Ahmed}\affiliation{\REG}  
\author{B.~Aljawrneh}\affiliation{\NCAT}  
\author{S.~Alsalmi}\affiliation{\KENT}\affiliation{\KSU} 
\author{R.~Ambrose}\affiliation{\REG} 
\author{W.~Armstrong}\affiliation{\TEMP} 
\author{A.~Asaturyan}\affiliation{\JLAB} 
\author{K.~Assumin-Gyimah}\affiliation{\MSU}   
\author{C.~Ayerbe Gayoso}\affiliation{\WM}\affiliation{\MSU}    
\author{A.~Bandari}\affiliation{\WM}   
\author{S.~Basnet} \affiliation{\REG}  
\author{V.~Berdnikov}\affiliation{\CUA}\affiliation{\JLAB}    
\author{H.~Bhatt}\affiliation{\MSU}   
\author{D.~Bhetuwal}\affiliation{\MSU}
\author{W.~U.~Boeglin}\affiliation{\FIU} 
\author{P.~Bosted}\affiliation{\WM} 
\author{E.~Brash}\affiliation{\CNU}    
\author{M.~H.~S.~Bukhari}\affiliation{\JAZ}  
\author{H.~Chen}\affiliation{\UVA}             
\author{J.~P.~Chen}\affiliation{\JLAB}           
\author{M.~Chen}\affiliation{\UVA}             
\author{M.~E.~Christy}\affiliation{\HU}\affiliation{\JLAB}          
\author{S.~Covrig~Dusa}\affiliation{\JLAB}           
\author{K.~Craycraft}\affiliation{\UTENN}        
\author{S.~Danagoulian}\affiliation{\NCAT}     
\author{D.~Day}\affiliation{\UVA}             
\author{M.~Diefenthaler}\affiliation{\JLAB}     
\author{M.~Dlamini}\affiliation{\OHIO}          
\author{J.~Dunne}\affiliation{\MSU}            
\author{B.~Duran}\affiliation{\TEMP}            
\author{D.~Dutta}\affiliation{\MSU}
\author{R.~Ent}\affiliation{\JLAB}  
\author{R.~Evans}\affiliation{\REG}            
\author{H.~Fenker}\affiliation{\JLAB}           
\author{N.~Fomin}\affiliation{\UTENN}            
\author{E.~Fuchey}\affiliation{\UCONN}           
\author{D.~Gaskell}\affiliation{\JLAB}          
\author{T.~N.~Gautam}\affiliation{\HU}         
\author{J.~O.~Hansen}\affiliation{\JLAB}           
\author{F.~Hauenstein}\affiliation{\ODU}\affiliation{\JLAB}       
\author{A.~V.~Hernandez}\affiliation{\CUA}       
\author{T.~Horn}\affiliation{\CUA}             
\author{G.~M.~Huber}\affiliation{\REG}       
\author{M.~K.~Jones}\affiliation{\JLAB}          
\author{S.~Joosten}\affiliation{\ANL}          
\author{M.~L.~Kabir}\affiliation{\MSU}
\author{C.~Keppel}\affiliation{\JLAB}           
\author{A.~Khanal}\affiliation{\FIU}           
\author{P.~M.~King}\affiliation{\OHIO}             
\author{E.~Kinney}\affiliation{\BOULDER}           
\author{M.~Kohl}\affiliation{\HU}              
\author{N.~Lashley-Colthirst}\affiliation{\HU}        
\author{S.~Li}\affiliation{\UNH}               
\author{W.~B.~Li}\affiliation{\WM}               
\author{A.~H.~Liyanage}\affiliation{\HU}         
\author{D.~Mack}\affiliation{\JLAB}              
\author{S.~Malace}\affiliation{\JLAB}           
\author{P.~Markowitz}\affiliation{\FIU}       
\author{J.~Matter}\affiliation{\UVA}           
\author{D.~Meekins}\affiliation{\JLAB}          
\author{R.~Michaels}\affiliation{\JLAB}         
\author{A.~Mkrtchyan}\affiliation{\YER}         
\author{H.~Mkrtchyan}\affiliation{\YER}        
\author{Z.~Moore}\affiliation{\JMU}
\author{S.J.~Nazeer}\affiliation{\HU}         
\author{S.~Nanda}\affiliation{\MSU}   
\author{G.~Niculescu}\affiliation{\JMU}        
\author{I.~Niculescu}\affiliation{\JMU}        
\author{D.~Nguyen}\affiliation{\UVA} 
\author{Nuruzzaman}\affiliation{\RUTG}       
\author{B.~Pandey}\affiliation{\HU}\affiliation{\VMI}           
\author{S.~Park}\affiliation{\SBU}             
\author{E.~Pooser}\affiliation{\JLAB}           
\author{A.~J.~R.~Puckett}\affiliation{\UCONN}         
\author{M.~Rehfuss}\affiliation{\TEMP}          
\author{J.~Reinhold}\affiliation{\FIU}         
\author{B.~Sawatzky}\affiliation{\JLAB}          
\author{G.~R.~Smith}\affiliation{\JLAB}             
\author{H.~Szumila-Vance}\affiliation{\JLAB}\affiliation{\FIU}
\author{A.~S.~Tadepalli}\affiliation{\RUTG}        
\author{V.~Tadevosyan}\affiliation{\YER}        
\author{R.~Trotta}\affiliation{\CUA}           
\author{S.~A.~Wood}\affiliation{\JLAB}            
\author{C.~Yero} \affiliation{\FIU}\affiliation{\CUA}  
\author{J.~Zhang}\affiliation{\SBU}        
\collaboration{for the Hall C Collaboration}
\noaffiliation

\begin{abstract}
Nucleon structure functions, as measured in lepton-nucleon scattering, have historically provided a critical observable in the study of partonic dynamics within the nucleon. However, at very large parton momenta it is both experimentally and theoretically challenging to extract parton distributions due to the probable onset of non-perturbative contributions and the unavailability of high-precision data at critical kinematics. Extraction of the neutron structure and the d-quark distribution have been further challenging because of the necessity of applying nuclear corrections when utilizing scattering data from a deuteron target to extract the free neutron structure.  However, a program of experiments has been carried out recently at the energy-upgraded Jefferson Lab electron accelerator aimed at significantly reducing the nuclear correction uncertainties on the d-quark distribution function at large partonic momentum.  This allows leveraging the vast body of deuterium data covering a large kinematic range to be utilized for d-quark parton distribution function extraction. In this paper, we present new data from experiment E12-10-002, carried out in Jefferson Lab Experimental Hall C, on the deuteron to proton cross--section ratio at large Bjorken-$x$.  These results significantly improve the precision of existing data and provide a first look at the expected impact on quark distributions extracted from parton distribution function fits.  
\end{abstract}

\maketitle

Measurements of the nucleon $F_2$ structure function in inelastic lepton-nucleon scattering and the kinematic evolution of $F_2$ occupy a prominent place in the historical development and testing of the theory of the strong interaction, Quantum Chromodynamics (QCD) \cite{RevModPhys.63.573,RevModPhys.63.597,RevModPhys.63.615}.  Such measurements have provided critical data in perturbative QCD (pQCD) fits used to extract quark and gluon distributions and in testing the universality of the pQCD evolution equations of these parton distribution functions (PDFs) \cite{accardi2010,alekhin01,alekhin03}.  While tremendous progress has been made in this endeavor over the last few decades, much is still left to be fully explored.  One such example is the longitudinal momentum distribution of the down quarks when the nucleon's momentum is predominantly carried by a single valence quark, or as $x$ $\rightarrow$ 1. Here $x$ is the Bjorken "scaling" variable which can be interpreted as the fractional momentum of the nucleon carried by the struck quark.   While there exists a number of effective theory predictions \cite{akp17,hybrid,alekhin01,alekhin03, cj15} for the ratio of the proton down to up quark distributions (d/u) at large $x$, additional experimental data are required to adequately test these.  
\begin{figure}
\begin{centering}
\includegraphics[trim=0 10 0 20,clip,width=\linewidth]{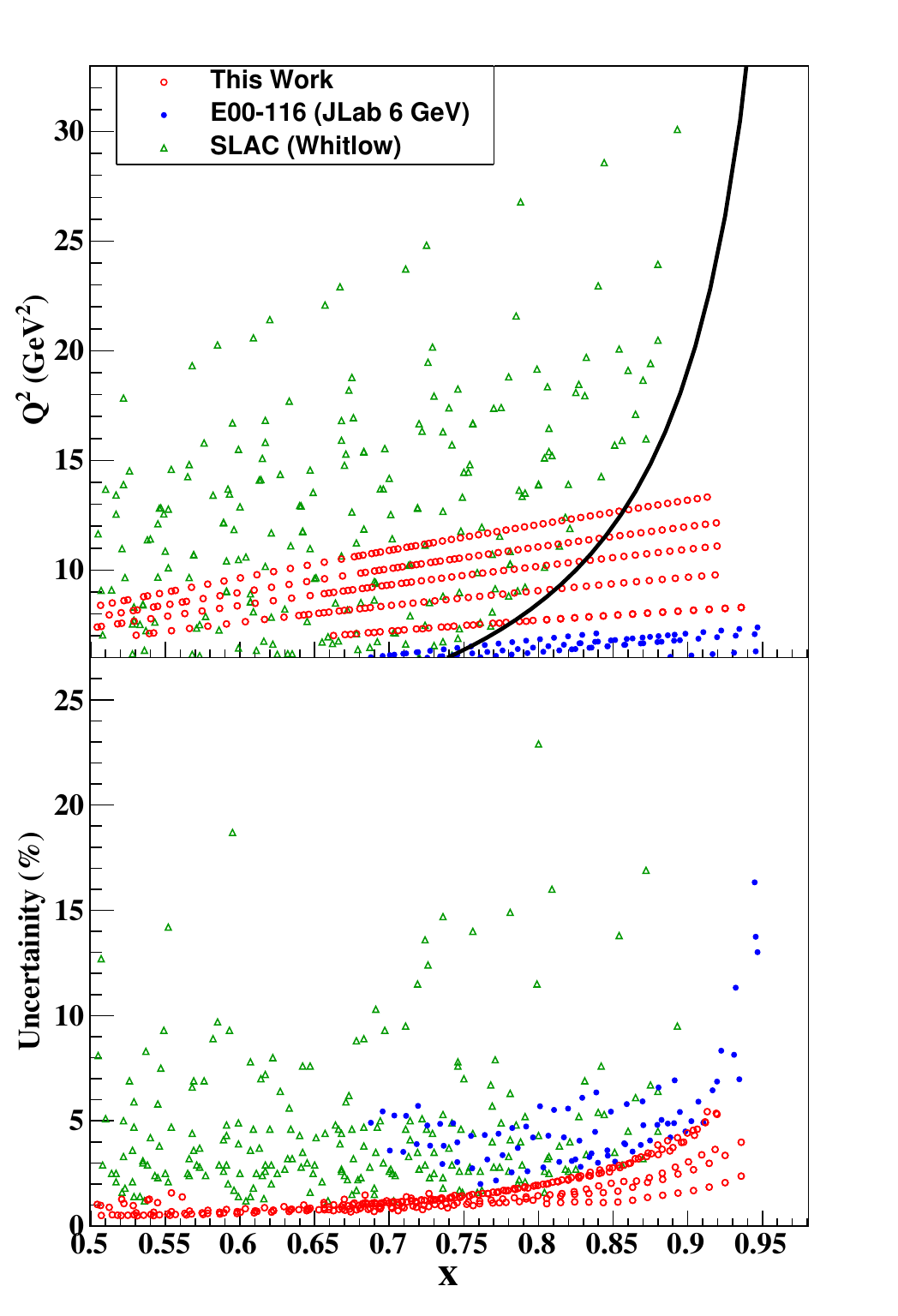}
\end{centering}
\caption{The top panel shows the high-x kinematic coverage of this work (red circles), compared with the Whitlow reanalysis \cite{Whitlow,Whitlow:1991} existing SLAC data (green triangles).  The solid blue circles are from JLab's 6 GeV experiment, E00-116 \cite{Malace:2006}.  Only data where $x > 0.5$ and $Q^2 > 6$ are shown.  The solid curve indicates $W^2=3$ GeV$^2$, where $W$ is the invariant mass of the produced hadronic system.  The statistical uncertainty of the deuteron to hydrogen cross--section ratio from these experiments are shown in the bottom panel.}
\label{fig:q2_vs_x}
\end{figure}
The last few years have seen the completion of three complementary experiments performed at Jefferson Lab utilizing the energy-upgraded CEBAF accelerator and aimed at extracting the neutron to proton $F_2$ ratio and providing access to d/u at large $x$.  The first of these was the MARATHON~\cite{marathonprl} experiment in Hall A, which measured ratios of the inclusive structure function $F_2$ from the A=3 mirror nuclei $^3$He and $^3$H, as well as from the deuteron and proton.  The second experiment was the BONuS12~\cite{bonusProposal} experiment in Hall B, which is a follow-up to the BONuS~\cite{bonus6,bonusEmc,bonusNeutron} experiment, but leveraging the doubling of the beam energy to 12~GeV to access larger $x$ without entering the region of the nucleon resonances.  Jefferson Lab (JLab) experiment E12-10-002 (this work) measured $H(e,e')$ and $D(e,e')$ inclusive cross--sections with the aim of extracting the hydrogen and deuterium $F_2$ structure functions at large $x$ and intermediate four-momentum transfer, $Q^2$.  The new high-precision data from this work, especially when coupled with new nuclear correction data from BONuS12 and MARATHON, will provide new insight into the up and down quark distributions within the nucleon.

The dataset was acquired in February--March of 2018 in Hall C. The experiment used the High Momentum Spectrometer (HMS), the SuperHMS (SHMS), and liquid cryogenic hydrogen and deuterium targets. The electron beam energy was 10.602~GeV and the current varied between 30 and 65~$\mu$A. The experiment served as one of the commissioning experiments for the new or upgraded Hall C equipment associated with the JLab 12~GeV energy upgrade. The data were acquired in ``scans'' at a fixed spectrometer angle by varying the central momentum setting and alternating between the 10 cm long hydrogen and deuterium targets. The results presented here stem from five different SHMS scans at (nominal) scattering angles of 21, 25,  29, 33, and 39 degrees. The central momentum varied between 1.3 and 5.1~GeV/c.  Additional scans were taken with the HMS at 21 and 59 degrees.  The 21 degree data were used as a cross-check between the well understood HMS and the newly constructed SHMS.  The 59 degree data are still being analyzed and are not presented here.  The kinematic coverage of this work, in $Q^2 $ and $x$ coordinates, is shown in Figure \ref{fig:q2_vs_x}, also displayed are the world data from SLAC (green triangles) and 6 GeV JLab (blue solid circles).  Prior to this work, the invariant mass region of $W^2<3$~GeV$^2$, (i.e. to the right of the solid curve), is poorly populated above a $Q^2$ of about 6 GeV$^2$.  The statistical uncertainties of this work, shown in the top panel of Fig.\ref{fig:q2_vs_x}, are a vast improvement over existing data. 

\begin{figure}
\begin{centering}
\includegraphics[trim=10 0 60 40 ,clip,width=\linewidth]{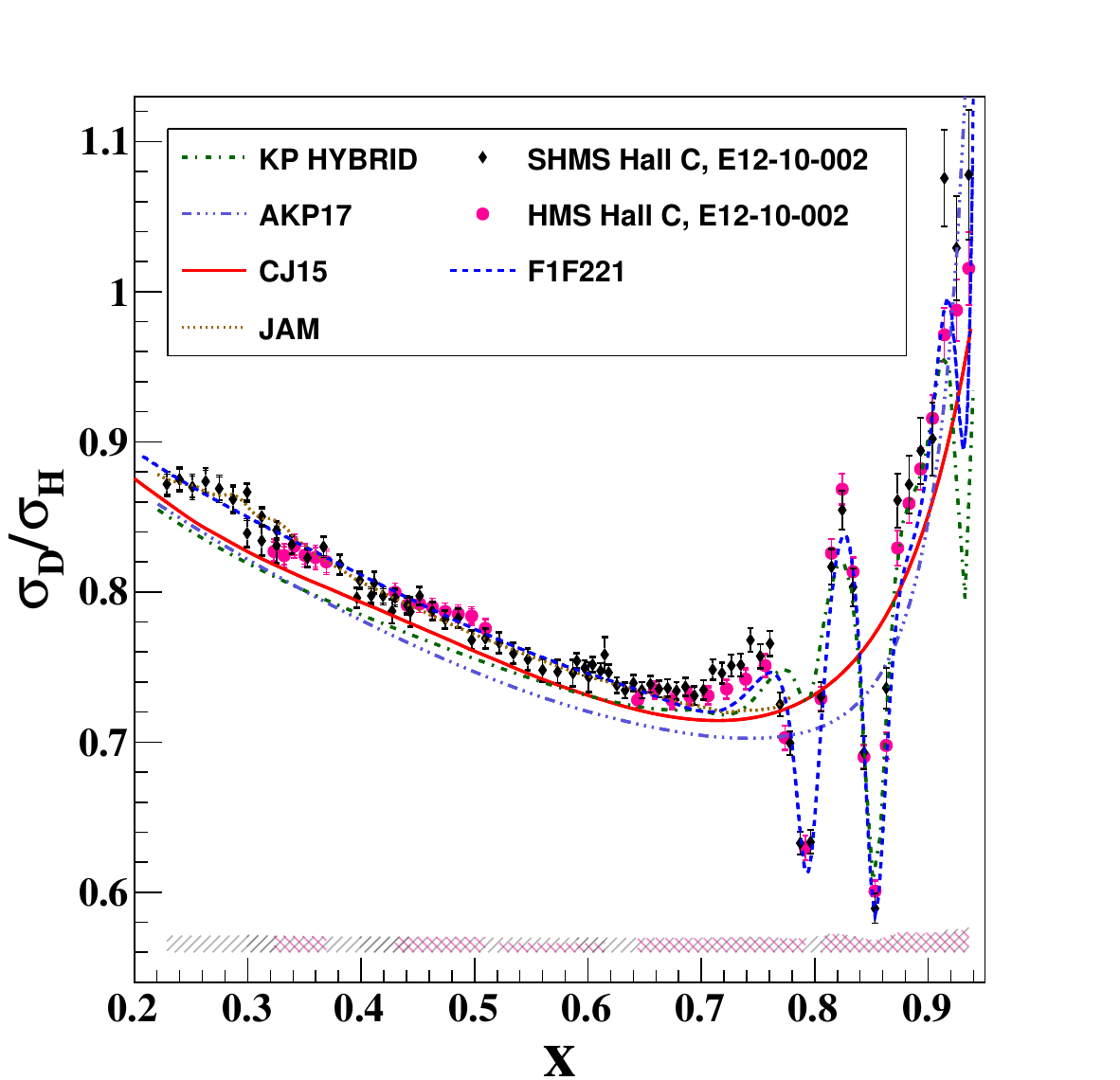}
\end{centering}
  \caption{The $\sigma_{D}/\sigma_H$ ratio as a function of $x$ for a spectrometer angle of 21 deg ($Q^2$ range from 3.39 to 8.25 GeV$^2$).  To first order,  the cross section ratio is equal to the $F_2$ structure function ratio. The error bars include uncorrelated systematic and statistical errors. The error bands include correlated systematic errors and an overall normalization uncertainty of 1.1\%(see Table \ref{table:error}.).  F1F221 (blue dashed line) is the model used in this analysis, the other curves are from different PDF fits (see text).  Good agreement is observed between the HMS and SHMS spectrometers. }
\label{fig:21deg}
\end{figure}

The SHMS was a new spectrometer installed in Hall C to take advantage of the energy upgrade of the CEBAF accelerator to 12 GeV. \cite{shms_nim,ct_long,ct_prl}.
Its magnetic layout consists of a horizontal bender, three quadrupoles, and a dipole ($HQ\bar{Q}QD$). The maximum momentum is 11.0 GeV/c, the typical momentum acceptance is -10\% to 22\% about the central momentum, and the solid angle is $\sim 4.0 $~msr. The standard detector package includes a gas Cherenkov detector (filled with 1 atm of CO$_2$)  and an electromagnetic calorimeter for particle identification (PID), two wire drift chambers for tracking and event reconstruction, and four hodoscope planes used in the event trigger.  An additional heavy gas Cherenkov was present in the detector package but not used in this analysis as it is primarily used for hadron identification. 

In the one-photon exchange approximation the differential cross--section for inclusive electron scattering can be written as:

\begin{equation}
\label{eq:xsect_R}
\frac{d^2 \sigma}{d \Omega d E^\prime} = \sigma_{\textrm{Mott}}\frac{2 M x F_2}{Q^2\varepsilon}\Big(\frac{1+\varepsilon R}{1 + R} \Big)
\end{equation}
Where $\sigma_{\textrm{Mott}}$ is the Mott cross--section, $M$ is the nucleon mass, $Q^2$ is the negative of the four-momentum transfer squared, $R$ is the ratio of the longitudinal to transverse photoabsorption cross--sections ($R = \sigma_L/\sigma_T$) and $\varepsilon$ is the virtual photon polarization.  The aim of this work is to obtain the $F_2^D/F_2^H$ structure function ratio, as it presents several advantages theoretically as well as experimentally. By reporting a quantity involving deuterium rather than the (``free'') neutron, we avoid choosing a particular prescription for treating nuclear effects, allowing theory groups active in this field to extract $F_2^\textrm{n}$ using their own nuclear corrections. Furthermore, the $\sigma_L/\sigma_T$ ratio is largely the same for hydrogen and deuterium \cite{E140X:1995ims}, thus, to first order, the $F_2^D/F_2^H$ ratio is the same as the cross--section ratio.  The experimental advantage of reporting a cross section ratio is that several corrections to the yield cancel (e.g. detector efficiencies) and multiple systematic errors are reduced such as the effective target length, deadtime corrections and spectrometer acceptance.
\begin{figure*}
\begin{centering}
\includegraphics[trim=0 0 0 0,clip,width=.9\textwidth]{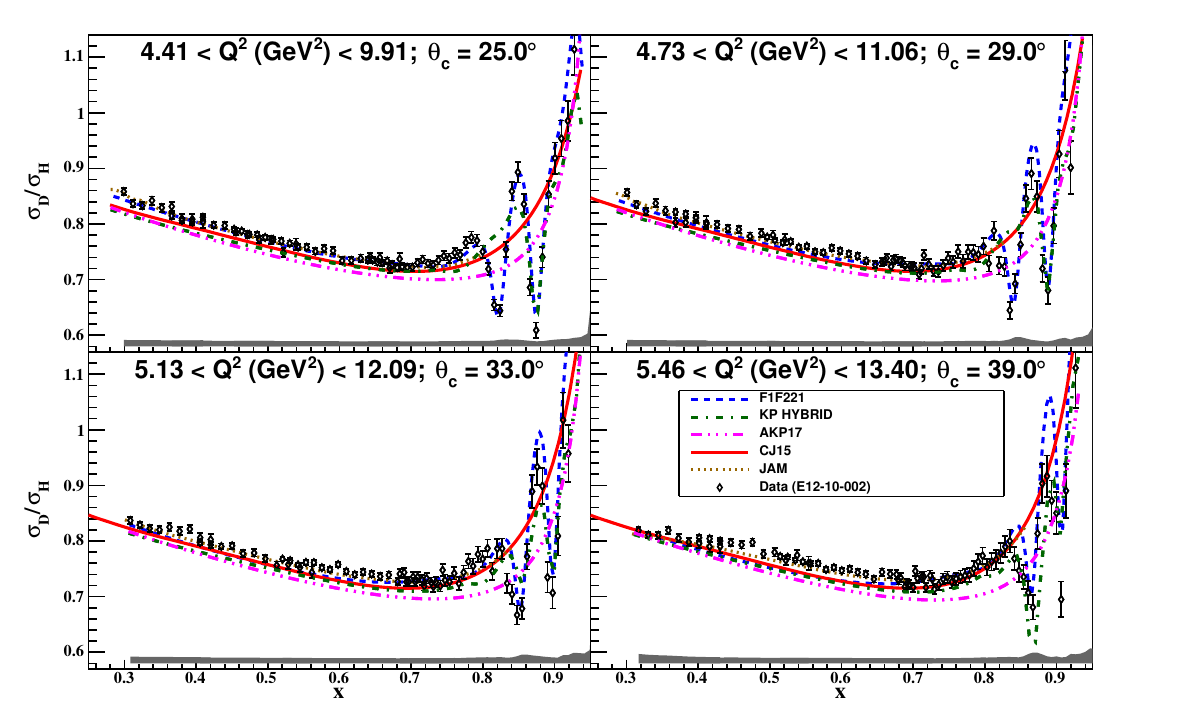}
\end{centering}
   \caption{The cross section ratio, $\sigma_D/\sigma_H$, as a function of $x$ for SHMS spectrometer angles of 25, 29, 33, and 39 deg.  To first order,  the cross section ratio is equal to the $F_2$ structure function ratio.  The $Q^2$ range of each setting is indicated in each panel.  }
\label{fig:4panel}
\end{figure*}

Experimentally, the deuteron to proton cross--section ratio,  $\left ( \sigma_D/\sigma_H\right )_{\textrm{Exp}}$, is obtained using the Monte Carlo ratio method \cite{halla_xsec} 
\begin{equation}
\label{xsect2}
\left(\frac{\sigma_D}{\sigma_H}\right )_{\textrm{Exp}} = \frac{\mathcal{R_D}}{\mathcal{R_H}}\left(\frac{\sigma_D}{\sigma_H}\right)_{\textrm{Model}}
\end{equation}
where $\mathcal{R}=Y_{\textrm{Data}}/Y_{\textrm{MC}}$, $Y_{\textrm{Data}}$is the efficiency and background corrected charge normalized electron yield, $Y_{\textrm{MC}}$ is the Monte Carlo yield obtained using a model cross--section that is radiated using the Mo and Tsai formalism \cite{Tsai, MoTsai}, and $\left ( \sigma_D/\sigma_H\right )_{\textrm{Model}}$ is the same model cross--section evaluated at the Born level.  The yields were binned in $W^2$, and then converted to $x$.  Electrons were selected by applying cuts to the gas Cherenkov and the energy deposited in the calorimeter normalized by the momentum of the track.  

Corrections to $Y_{\textrm{Data}}$, along with their magnitudes relative to $\left ( \sigma_D/\sigma_H\right )_{\textrm{Exp}}$ in the inelastic region ($W^2 > 3$~GeV$^2$), include contributions from pion contamination (0.907-1.000), deadtime (0.682-1.342), target density (0.989-0.998), tracking efficiency (0.987-1.000), trigger efficiency (1.000-1.004), and backgrounds from the target cell walls (1.033-1.060).  Pions that pass the electron PID cuts were removed using a parameterization of the pion contamination as a function of the scattered electron energy, $E'$ \cite{sun}.  The computer deadtime was found by comparing the number of triggers recorded in scalers to the number found in the datastream.  The electronic deadtime, due to events being lost at the trigger logic level, was measured by injecting a pulser of known frequency at the start of the trigger logic chain.  These pulser events, identifiable via TDC information, were compared with the number of events recorded in the scalers.  Tracking efficiency was calculated by taking the ratio of events with detected tracks to the number of events that passed PID, fiducial and timing cuts.  The trigger for this experiment required signals in 3 of the 4 hodoscope layers and a signal in either the gas Cherenkov or calorimeter.  The trigger efficiency was $>99\%$ and determined by calculating the efficiency of the individual hodoscope planes.  Backgrounds from the aluminum cell walls were subtracted from the cryogenic targets by utilizing ``dummy'' data taken on two aluminum targets placed at the same location as the cryogenic entrance and exit windows.  A target density correction was applied to account for a local change in density due to heating from the electron beam.  A series of dedicated measurements at various currents up to 80 $\mu$A were performed and the charge normalized yields were plotted vs beam current.  The density reduction for the hydrogen (deuterium) target was $2.55 \pm 0.74\frac{~\%}{100~ \mu A}$~$(3.09\pm0.84~\frac{\%}{100~\mu A})$.  For further details of the analysis see \cite{fernando,aruni, sun, deb, abishek, casey}. 

\begin{figure}
\begin{centering}
\includegraphics[trim=20 30 40 60,clip,width=\linewidth]{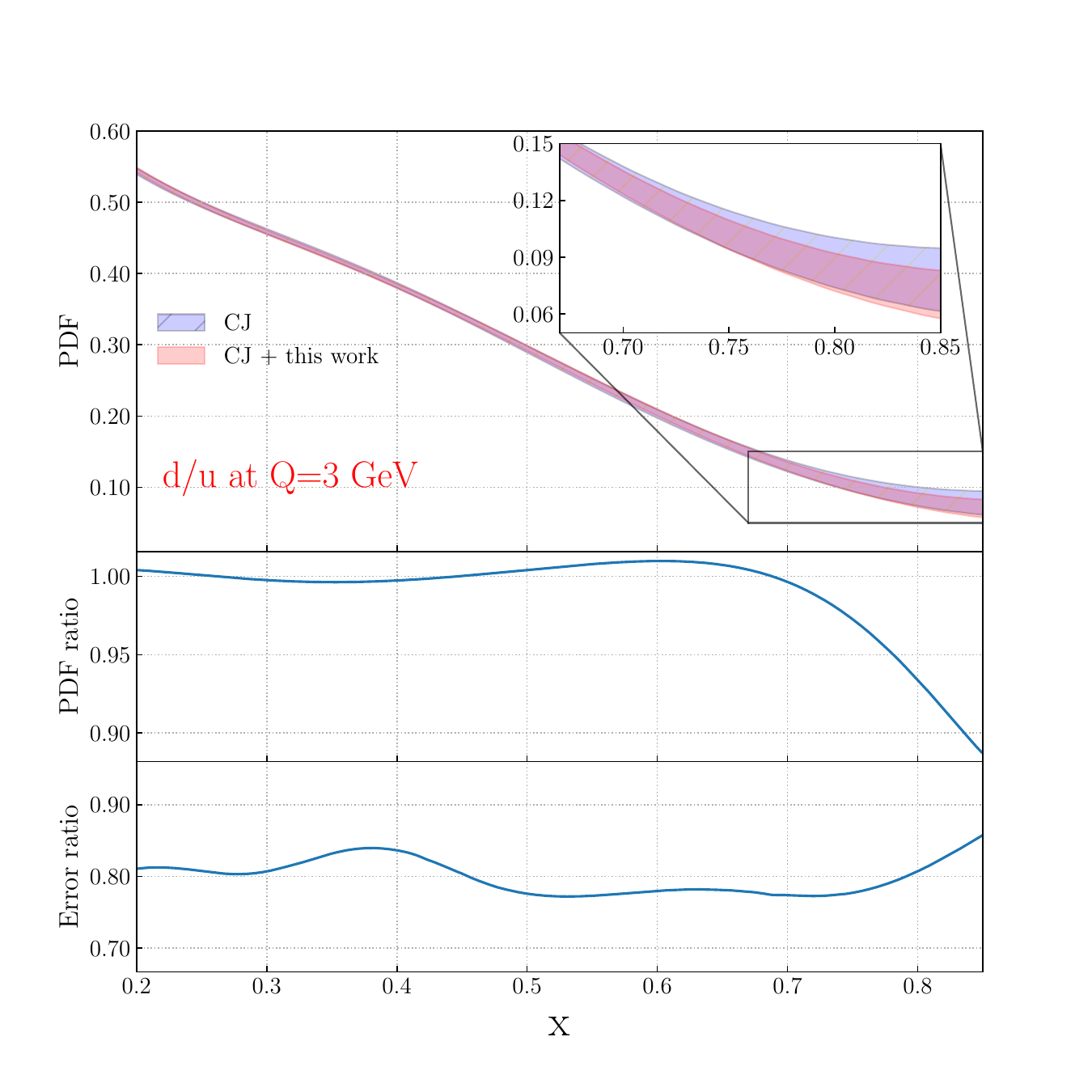}
\caption{Top: The d/u ratio for the proton from CJ15 (blue) and when this work is included to the dataset used in CJ15 (red).  Middle: The relative change in the d/u PDF central value, the shift at $x > 0.7$ is due to the previous lack of deuterium data at high-x.  Bottom: The reduction in the d/u PDF relative uncertainty.  The inclusion of the data from this work results in a roughly $20\%$ reduction in the uncertainty. While a typical cut of $W^2>$  3.0 GeV$^{2}$ is used to eliminate the resonance region in the CJ15 framework, a cut of $W^2>$  3.5 GeV$^{2}$ was applied to the new dataset.}
\end{centering}
\label{cj}
\end{figure}
Electrons produced by charge symmetric backgrounds, mainly from neutral pion production (e.g. $\pi^0 \rightarrow \gamma\gamma^{\ast} \rightarrow \gamma e^+e^-)$, in which the photon decays into a positron and an electron were included in the Monte Carlo yield.    This background was measured by reversing the spectrometers' magnet polarity to measure the positron yield for both hydrogen and deuterium targets.  The background was parameterized with a two parameter fit as a function of $E'$.  Due to beam time constraints, positron data was acquired for only three of the five angular settings.  To circumvent this limitation, the positron yield was parameterized as described in \cite{Vahe}.  The parameterization was then used to extrapolate the positron yield to the kinematic settings where measurements were not available. For $x > 0.6$, the background contribution to the measured cross-section was less than 1\% and rose to 30\% with decreasing $x$ at the 39 degree angle setting. Additionally, the measured positron yield per nucleon was identical for both targets.

\begin{centering}
 \begin{table}
     \centering
          \begin{tabular}{| l |l|l|}
            \hline
        Error & Pt. to Pt (\%) & Correlated (\%) \\
            \hline
         Statistical & $0.5 - 5.4 (2.9)$ &\\
         Charge & $0.1 - 0.6$ &\\
         Target Density & $0.0 - 0.2 $ & $1.1$ \\
         Livetime && $0.0 - 1.0$ \\
         Model Dependence && $0.0 - 2.6 (1.2)$ \\
         Charge Sym. Background&&$0.0 - 1.4$ \\
         Acceptance && $0.0 - 0.6 (0.3)$ \\
         Kinematic && $0.0 - 0.4 $ \\
         Radiative Corrections && $0.5 - 0.7 (0.6)$ \\
         Pion Contamination && $0.1- 0.3$ \\
         Cherenkov Efficiency && $0.1$ \\
         \hline
         Total  & $0.6 - 5.4 (2.9) $& $1.2 - 2.9 (2.1)$ \\
          \hline
           \hline
  \end{tabular}
   \caption{\label{table:error}The error budget for the cross--section ratio  $\sigma_{D}/\sigma_{H}$.  The error after a cut of $W^2>3$ GeV$^2$ is shown in parenthesis, which is a typical cut applied to eliminate the resonance region while performing PDF fits.}
\end{table}
\end{centering}

The uncertainties in the deuterium to hydrogen cross--section ratio $\sigma_{D}/\sigma_{H}$, shown in Table \ref{table:error}, are divided into two categories, uncorrelated point-to-point and correlated (see \cite{supp} for more details).  An overall normalization uncertainty of 1.1\% due to uncertainty in the target density is included in the correlated error.  The target density error includes uncertainties from the target temperature and pressure, measured length, thermal contraction, the equation of state used to calculate the density, and the target boiling correction.  Additional point-to-point errors for target density are included to account for runs where the boiling correction was far from the average due to higher or lower beam currents.
A cross--section model dependence error was assessed by altering the model fit parameters, thereby changing $\left ( \sigma_D/\sigma_H\right )_{\textrm{Model}}$ by as large as 10\%. The change in $\left ( \sigma_D/\sigma_H\right )_{\textrm{Exp}}$  was parameterized as a function of $W^2$ and used as the uncertainty.
The most significant effects were observed at higher $x$ values, where the resonance region causes rapid changes in the cross-section. Binning in $W^2$, as opposed to $E'$ or $x$, was found to reduce this uncertainty due to the location of the resonance peaks being fixed.  Errors from the radiative corrections include a contribution from both the model and the method.  The model dependence was determined by scaling the various quasi-elastic contributions to the model.  The error associated with the method (0.5\%) was taken from \cite{dasu}.  A kinematic uncertainty was determined with Monte Carlo by individually varying the beam energy and central momentum of the spectrometer by $\pm 0.1\%$ and also by varying the spectrometer angle by $\pm 0.25~$mr.  

The results of this analysis are summarized in Fig.~\ref{fig:21deg} and Fig.~\ref{fig:4panel}, the numerical values can be found in the tables included in the Supplemental Material \cite{supp}. The $\sigma_D/\sigma_H$ ratio is shown as a function of $x$ for each of the SHMS spectrometer angles.  The curves shown are predictions for $F_2^D/F_2^H$ obtained using four available models evaluated at the same kinematics as the data: CJ15\cite{cj15} (solid red line), AKP17\cite{akp17} (dot-dot-dot-dashed violet line), KP Hybrid\cite{hybrid} (dot-dashed line) and JAM  \cite{jam1,jam2} (dotted brown line).  The model used to extract the cross--section is F1F221 (dashed blue line) which is an improved fit to world data \cite{f1f209} . None of the models shown includes the data from this analysis.  

As a representative example of the theoretical  impact of this experimental result, the dataset from this work \cite{supp2} was evaluated with the CJ15 QCD analysis \cite{cj15} framework, which deploys state-of-the-art deuteron nuclear corrections.  This next-to-leading-order analysis utilizes data from a variety of high energy scattering processes, including deep inelastic scattering, weak boson and jet production, and Drell-Yan.  A fitted normalization factor of $-2.1\%$ was determined in order for the data set to agree with the CJ model \cite{privateLi}, slightly larger than the total $x$ dependent correlated error of 1.3$\%$-2.1$\%$ shown in Table \ref{table:error}.  Furthermore, this experiment ran in parallel with E12-10-007 (a measurement of the EMC effect) which observed a 2.0\% normalization difference with previous EMC measurements\cite{emc12}, the direction of this normalization difference is consistent with that found in the CJ15 study.  The fitted PDFs with and without this experiment were compared at the central value as well as the size of uncertainties. For consistency, the error band for each fit was calculated at 90\% confidence level \cite{cj2024}.  Fig. 4 depicts the d/u ratio, a fundamental quantity and testing ground for multiple (p)QCD predictions regarding nucleon structure.  The fitted d/u PDF before and after inclusion of this data is shown, where the significant reduction in the uncertainties demonstrates the importance of high precision data in PDF extractions. Not only did the inclusion of this work reduce the relative error by approximately 20\% across the entire $x$ range, but it also shifted the d/u central value at large x by as much as 10\%.  Furthermore, this data provides additional constraints on the parameters used in higher twist corrections, the individual d and u quark distributions, and the target mass corrections used in these fits.  

It should be noted that, on average, the ratio of deuterium to hydrogen cross--section from this work and MARATHON\cite{marathon1} differs by as much as 4.3\% or 2$\sigma$ where the datasets overlap in the $x$ range of 0.2 - 0.3, with this work being above the MARATHON result. However, in a recent QCD analysis\cite{wally} a normalization factor of $+1.9\%$ was required to get the MARATHON d / p data to agree with the existing data.  If this normalization is applied, together with the normalization factor found in the above CJ15 study, the two datasets agree within 0.3$\%$.  All the aforementioned data agree with the previously available SLAC data, which have large uncertainties \cite{Whitlow}. 

In summary, high-precision inclusive measurements on hydrogen and deuterium were performed for $Q^2$ from 3.4 to 13.4 GeV$^{2}$ and $x$ from 0.3 to 0.93. Inclusion of this dataset with one PDF fitting framework resulted in a 20\% uncertainty reduction in the proton d/u, and when combined with the MARATHON and BoNUS12 results, will have a significant impact on understanding PDFs in the valence regime.  It can be used, moreover, for quark-hadron duality studies, spin-flavor symmetry breaking, and constraints on nuclear corrections. The results of this work will also be beneficial for the future EIC and DUNE\cite{dune} experimental programs.  DUNE will benefit from higher precision neutrino oscillation Monte Carlos and the EIC can have higher precision cross-section predictions, where structure function information at large $x$ feeds down through pertubative $Q^2$ evolution to lower $x$ and higher values of $Q^2$.
Moreover, the combination of precision PDFs at high $x$ and low $Q^2$ with low $x$ and high $Q$ will allow for critical studies of the onset of non-pertubative effects and inter-nucleon bindings. 

This material is based upon work supported by the U.S. Department of Energy, Office of Science, Office of Nuclear Physics under contracts DE-AC05-06OR23177, DE-AC02-05CH11231, DE-SC0013615, and DE-FE02- 96ER40950, and by the National Science Foundation (NSF) Grant No. PHY-1913257, No. PHY-1914034 and No. PHY-2013002.


\bibliographystyle{apsrev4-2}
\bibliography{e12_10_002}

\end{document}